\documentclass[rnote]{aa} 
\usepackage{graphicx}
\usepackage{lscape}
\usepackage{subfigure}
\usepackage{natbib}
\bibpunct{(}{)}{;}{a}{}{,} 
\usepackage{stfloats}
\usepackage{txfonts}
\begin{document}
\title{\textbf{The spectral type of} CHS\,7797 - an intriguing very low mass periodic variable in the Orion Nebula Cluster.}

 	\author{Mar\'{i}a V. Rodr\'{i}guez-Ledesma
          \inst{1,}\inst{2},         
          Reinhard Mundt\inst{1},
Olga Pintado
\inst{3},
Steve Boudreault
\inst{4,}\inst{5},
Frederic Hessman
\inst{2}
\and
William Herbst
          \inst{6}
	 }
\institute{Max-Planck-Institut f\"ur Astronomie (MPIA), K\"onigstuhl 17, D-69117 Heidelberg, Germany\\
              \email{vicrodriguez@mpia.de}
	     \and
Institut f\"ur Astrophysik, Georg-August-Universit\"at, Friedrich-Hund-Platz 1, D-37077 G\"ottingen, Germany\\
\and
Instituto Superior de Corellaci\'{o}n Geol\'{o}gica, CONICET, Miguel Lillo 205, 4000, San Miguel de Tucum\'{a}n, Argentina\\
\and
Instituto de Astrof\'{i}sica de Canarias (IAC), Calle V\'{i}a L\'{a}ctea s/n, E-38200 La Laguna, Tenerife, Spain\\ 
\and
Departamento de Astrof\'{i}sica, Universidad de La Laguna (ULL),
E-38205 La Laguna, Tenerife, Spain\\
\and
Astronomy Department, Wesleyan University, Middletown, CT 06459 USA }


 
  \abstract
{}
{We present the spectroscopic characterization of the unusual high-amplitude very low mass pre-main-sequence periodic variable CHS\,7797.
}
{This study is based on optical medium-resolution (R$=$2200) spectroscopy in the $6450\,-\,8600\,$\AA$\,$ range, carried out with GMOS-GEMINI\,-S in March 2011. Observations of CHS\,7797 have been carried out at two distinct phases of the 17.8\,d period, namely at maximum ($I$\,$\approx$\,17.4\,mag) and four days before maximum ($I$\,$\approx$\,18.5\,mag). Four different spectral indices were used for the spectral classification at these two phases, all of them well-suited for spectral classification of young and obscured late M dwarfs. In addition, the gravity-sensitive Na\,I (8183/8195\,$\AA$) and K\,I (7665/7699\,$\AA$) doublet lines were used to confirm the young age of CHS\,7797.
  }
{From the spectrum obtained at maximum light we derived a spectral type (SpT) of M\,6.05\,$\pm$\,0.25, while for the spectrum taken four days before maximum the derived SpT is M\,5.75\,$\pm$\,0.25. The derived SpTs confirm that CHS\,7797 has a mass in the stellar-substellar boundary mass range. In addition, the small differences in the derived SpTs at the two observed phases may provide indirect hints that CHS\,7797 is a binary system of similar mass components surrounded by a tilted circumbinary disk, a system similar to KH\,15D.
}
{}
\keywords{stars: low-mass, pre-main sequence, brown dwarf - stars: variable, binary - technique: spectroscopic }
\titlerunning{CHS\,7797: Spectral type determination }
\authorrunning{Rodr\'iguez-Ledesma et al.} 
\maketitle
%
\section{Introduction}

In the past years, the interest of studying brown dwarfs (BDs) has strongly increased. This increase in the number of BD studies is tightly connected with the advances in technology. Particularly, the availability of large telescopes (8\,m to 10\,m class telescopes) and the development of larger imaging detectors, both in the optical and NIR/IR wavelengths, where crucial for detecting and characterising an increasing number of these cool and faint objects. Scientifically, BDs are interesting objects since they fill the gap between stars and planets and may provide constraints on understanding the origins of these astronomical objects. Particularly interesting are the studies involving very young BDs that are only a few Myr old \citep[e.g.][]{Reiners2009,Lodieu2009}. Since these young very low mass (VLM) objects are hotter than their older and cooler counterparts, they are easier to detect. Nevertheless, in star-forming regions where most objects are still embedded in the gas and dust out of which they have been formed, the detection of these faint objects can be more difficult. 

Several surveys, aiming to characterise the lower mass regime in young star-forming regions, succeeded in finding young BDs
with spectral types (SpT) extending into the L dwarf regime \citep[e.g.][]{Lodieu2011,Riddick2007a,Luhman2004}. In addition, photometric monitoring programs, mostly aimed at studying variability and rotation evolution, have also played an important role in the detection of many VLM stars and even brown dwarf candidates \citep[e.g.][]{Scholz2004,Rodriguez-Ledesma2009}, some of which were subsequently confirmed spectroscopically as BDs. 

Spectral classification of young BDs and VLM stars are mostly based on measurements of narrow-band optical and near-infrared (NIR) spectral indices. These indices are defined to quantify the relative strengths of certain atomic and molecular features that are clearly correlated with SpT. \cite{Riddick2007a} performed a detailed investigation of optical spectral indices suitable for the spectral classification of young and embedded BDs in the Orion Nebula Cluster (ONC). They concluded that some of the commonly used indices in field objects might not be ideal for the classification of young BDs because of their dependence on reddening (e.g. the pseudo-continuum indices) or because the measured wavelength regions contain typical emission lines from the active young object and/or the nebular environment (e.g. the TiO-5 index). Spectral classification can also be done by comparing the observed spectra with model spectra, although this approach has the disadvantage to rely on the accuracy of the model spectra, and it is well known that the models at young ages suffer from large uncertainties mainly due to the arbitrary selection of the initial conditions at the pre-main sequence phase\citep[e.g.][]{Baraffe2003}.\\
\begin{table*}[bp]
\centering
\renewcommand{\arraystretch}{1.6}  
\caption{Definition of the spectral indices used for the spectral type determination of CHS\,7797. The spectral indices are computed from the average flux over the defined wavelength ranges.}
\label{tab1}
\addtolength{\tabcolsep}{0.0pt}
\resizebox{14cm}{!}{
\begin{tabular}{c|c|c} 
\hline\hline             
Spectral Index & Definition$^{1}$  &Reference\\
\hline
TiO 8465&[8405 - 8425]/[8450 - 8470]&a,b\\
VO 2&[7920 - 7960]/[8130 - 8150] & a,c\\
VO 7445 & 0.5625[7350-7400]+0.4375[7510-7560]]/[7420 - 7470] &a,d\\
c81 & [8115-8165]/([7865-7915]+[8490-8540]) & a,e\\
\hline
\end{tabular}}
\renewcommand{\footnoterule}{} 

\begin{flushleft}
\small
 $^{1}$ Wavelength ranges are given in $\AA$.\\
References:\\
a) \cite{Riddick2007a}, b) \cite{Slesnick2006}, c) \cite{Lepine2003}, d) \cite{Kirkpatrick1995}, e) \cite{Stauffer1999}.
\end{flushleft}
\end{table*}
As mentioned above, extensive photometric monitoring programs have also been valuable in identifying interesting young VLM objects. This is also the case for CHS\,7797, an unusual high-amplitude 17.8\,d periodic variable in the ONC \citep{Rodriguez-Ledesma2012a}. As described in detail by \cite{Rodriguez-Ledesma2012a}, CHS\,7797 has a variation amplitude of $\approx$\,1.7 mag in the $R$, $I$, and z$\arcmin$ bands, which decreases only slightly at longer wavelengths (data available up to 4.5\,$\mu$m). The measured period over the six years (2004 to 2010) of monitoring in the $I$ band is $17.786\,\pm\,0.03$\,d with a very low false-alarm probability ($FAP\,=\,1\times10^{-15}\%$). The star is faint during $\approx$\,2/3 of the period and the shape of the phased light-curves for the seven different observing seasons shows minor changes and small-amplitude variations from one year to the other. 
Interestingly, there are no significant colour-flux correlations for $\lambda$\,$\lesssim$\,2\,$\mu$m. However, the object becomes redder when fainter at $\lambda$\,$\gtrsim$\,2\,$\mu$m, which indicates that CHS\,7797 is occulted by circumstellar matter in which grains have grown from typical 0.1\,$\mu$m to $\approx$\,1-2\,$\mu$m sizes. \cite{Rodriguez-Ledesma2012a} proposed two scenarios to explain the periodic variability observed over many years. In both scenarios the \emph{internal clock} for the variability is orbital motion in combination with extinction by a surrounding disc. In the first scenario CHS\,7797 is a single object surrounded by a planet/protoplanet with a period of 17.8 d, which gives rise to a disturbed and elongated structure within the circumstellar disc. In the second scenario CHS\,7797 is a 17.8\,d binary system similar to KH\,15D \citep[e.g.][]{Hamilton2001,Herbst2008}, in which an inclined circumbinary disc is responsible for the variability, and in which one component of the binary system is seen through the outer and less obscuring part of the disc \citep{Johnson2004,Winn2004,Winn2006,Chiang2004}. The photometric information of CHS\,7797 alone, although very rich, is not enough to characterise the system, and the spectral classification is then crucial to confirm its very low mass, its young age, and its possible binary character.

Other known young high-amplitude periodic variables, such as KH\,15D ,YLW\,16A, and WL\,4 in $\rho$\,Oph \citep{Plavchan2010,Plavchan2008}, and V718\,Per in IC 348 \citep{Grinin2008} were first identified by intensive variability studies and subsequently characterised spectroscopically. Among all these objects, CHS\,7797 is certainly by far the least massive one and might even be a brown dwarf binary system. 

In this research note we present the spectral classification of CHS\,7797. Spectroscopic observations are described in Section\,2. The spectral classification method and results are presented in Section\,3 and 4, while conclusions are drawn in Section\,5.

\section{Observations and data reduction}

Spectroscopic observations of CHS\,7797 have been carried out with the Gemini Multi-Object Spectrograph (GMOS) at the 8.1\,m GEMINI-S telescope at Cerro Pach\'{o}n, Chile (program GS-2011A-Q-8) during the nights of March 9 and 13, 2011. Long-slit observations have been performed with the R831/750 grism and a slit width of 1\,$\arcsec$, which provides an spectral resolution R\,$=$\,2200 and a nominal dispersion of 0.34\,$\AA$\,$/pix$ in the observed wavelength range of 6450-8600\,$\AA$. The selected spectral region allows for spectral type determination because many molecular absorption bands characteristic for M dwarfs can be measured (e.g. the TiO and VO absorption band strengths). Spectra of CHS\,7797 have been taken at two different phases of the 17.78\,d period \citep{Rodriguez-Ledesma2012a}, namely at maximum brightness (i.e. phase 0, $I$\,$\approx$\,17.4\,mag) and four days before maximum (i.e. phase $\approx$\,0.75, $I$\,$\approx$\,18.5\,mag). It is worth noting that due to the steep rise towards maximum light, the observations performed four days before maximum are close to the minimum brightness level \citep[see Fig.\,4 in][]{Rodriguez-Ledesma2012a}\footnote{We note that to observe a full amplitude variation in the spectra, observations should also have been taken at minimum brightness, i.e. phase 0.5.}. In each night, three spectra of CHS\,7797 were taken with an exposure time of 1800\,s each. The average S/N ratio (around $\lambda$\,$=$\,7500$\AA$) of the combined spectra taken at maximum brightness is $\approx$\,25, while the S/N ratio of the spectra taken at phase 0.75 is $\approx$\,15. Observations at the two distinct phases were performed under similar weather conditions and airmasses (i.e. airmass range from first to last spectrum taken during the two observing nights is 1.1\,-\,1.6). Seeing during observations varied between 0.6 and 0.8$\arcsec$.\\
For the data reduction we followed standard procedures using the gemini.gmos IRAF package. The three spectra were reduced separately and then combined for the final 1D-spectral extraction. The spectrophotometric standard star LTT215 was observed in the night of March 13 together with and
taken under the same conditions as the observations of CHS\,7797, to allow for a proper flux calibration.\\
The spectra of CHS\,7797 show strong emission of H$\alpha$ and forbidden emission lines like [SII] (6716/31\,$\AA$) and [NII] (6548/6583\,$\AA$). Although it is hard to separate the nebular from the intrinsic stellar emission, our analysis suggests that CHS\,7797 is a strong H$\alpha$ emitter (Rodriguez-Ledesma et al. 2012, in prep.).\\
Fig.\,\ref{Fig_spectra1} shows the spectra of CHS\,7797 at the two phases observed, i.e. maximum and four days before maximum. The spectra were normalised to the flux at 7500\,$\AA$ and displaced by a constant for clarity. Fig.\,\ref{Fig_spectra1} also shows the difference between the two spectra, showing minor differences between the spectra at the two observed phases for $\lambda$\,$\approx$\,8100\,-\,8400$\AA$.

\section{Spectral indices and SpT determination}

To determine the spectral type of CHS\,7797 we used a sample of narrow-band spectral indices specifically selected to measure the change of the strength of molecular absorption features with spectral type. \cite{Riddick2007a} investigated several temperature-sensitive molecular spectral indices that are suitable for an accurate spectral classification of young and embedded low-mass objects. The spectra of M dwarfs are dominated by TiO and VO molecular absorption bands, which are highly sensitive to temperature, i.e. the strength of these bands continuously increases with decreasing temperature until spectral types M7 and M9 for the TiO and VO bands, respectively. As pointed out by \cite{Riddick2007a}, this method provides precise spectral classification without the need of a direct comparison with other spectra.  
\begin{table}[!]
\centering
\renewcommand{\arraystretch}{1.3}  
\caption{Derived spectral index values and spectral types$^1$ for CHS\,7797 at the two observed phases, A (maximum) and B (four days before maximum).}
\label{tab2}
\addtolength{\tabcolsep}{0.0pt}
\resizebox{8.5cm}{!}{
\begin{tabular}{c|c|c|c|c} 
\hline\hline             
Spectral Index&Value&SpT&Value&SpT\\
&phase\,A&phase\,A&phase\,B&phase\,B\\
\hline
TiO 8465& 1.70 &M6.1& 1.65 & M6.0\\
VO 2& 0.52&M5.8& 0.56&M5.4 \\

VO 7445 & 1.04&M6.1&1.02&M5.8\\
c81 & 0.91 & M6.1&0.90&M5.8\\
\hline
\end{tabular}}
\renewcommand{\footnoterule}{} 
\begin{flushleft}
\small
$^1$\, SpT errors are 0.25 subclass.\\
\end{flushleft}
\end{table}
\begin{figure}
\centering
\includegraphics[width=8.2cm,height=7.3cm]{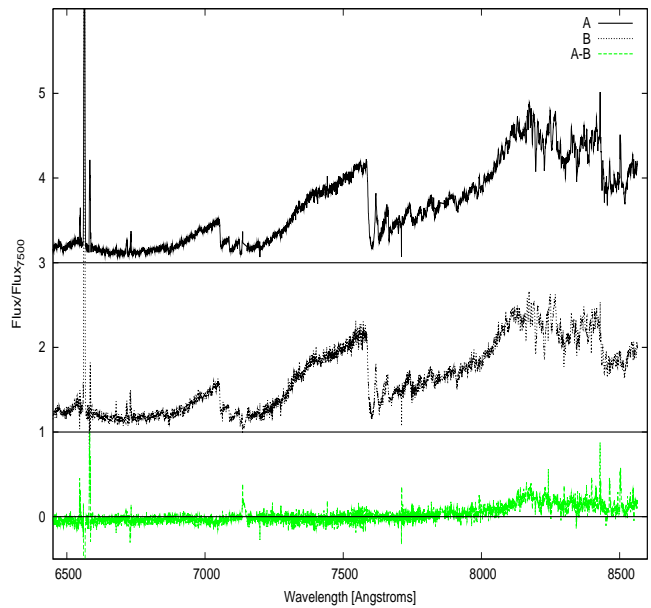}\\
\vspace{1cm}
\caption{GMOS spectra of CHS\,7797 at the two observed phases, i.e. maximum (A) and four days before maximum (B). The result of the subtraction of both spectra (A-B) is displayed in the bottom pannel. The zero-flux levels are indicated by horizontal lines. The spectra were normalised using the flux at 7500\,$\AA$ and displaced vertically by 1 and 3 units for the B and A phases, respectively. The spectra at these two phases are very similar, although an excess in the red side of the spectrum ($\lambda$\,$>$\,8000\,$\AA$) is evident at the maximum light phase. }
\label{Fig_spectra1}
\end{figure}
We used a set of four spectral indices for the spectral classification of CHS\,7797: TiO\,8465, VO-2, VO\,7445, and c81 as defined in Table\ref{tab1}. These molecular indices have the advantage to be almost unaffected by reddening and therefore are preferred for spectral classification of young embedded objects to the sometimes used pseudo-continuum indices (PC) \citep[e.g.][]{Martin1999}. A large number of indices have previously been defined in the wavelength range 6500-8500\,$\AA$, but many of them are not suitable for the spectral classification of CHS\,7797 because (1) they include regions of strong nebular emission lines (e.g. TiO-5 index), (2) they have long wavelength baselines and therefore are highly sensitive to reddening (e.g. PC3, PC4 indices), or (3) they include wavelength regions where the chip-gaps of the GMOS detector are located (e.g. TiO-5 index). It is also important to use an index-SpT relation calibrated for young objects, since the differences in surface gravity affect the spectral classification. \cite{Riddick2007a} calibrated 19 suitable spectral indices for a sample of 1-2\,Myr old M dwarfs in Taurus and Chameleon\,I \citep{Briceno2002, Luhman2004}. We rely on their detailed study and used the spectral index versus spectral type relations derived in their work. We present in Table\,\ref{tab2} the results obtained for the four suitable spectral indices computed at the two observed phases of CHS\,7797. As can be seen in Fig.\,\ref{Fig_spectra1} and also from the values presented in Table\,\ref{tab2} the two spectra are similar, but not identical. 

\begin{figure}
\centering
\includegraphics[width=8.2cm,height=7.3cm]{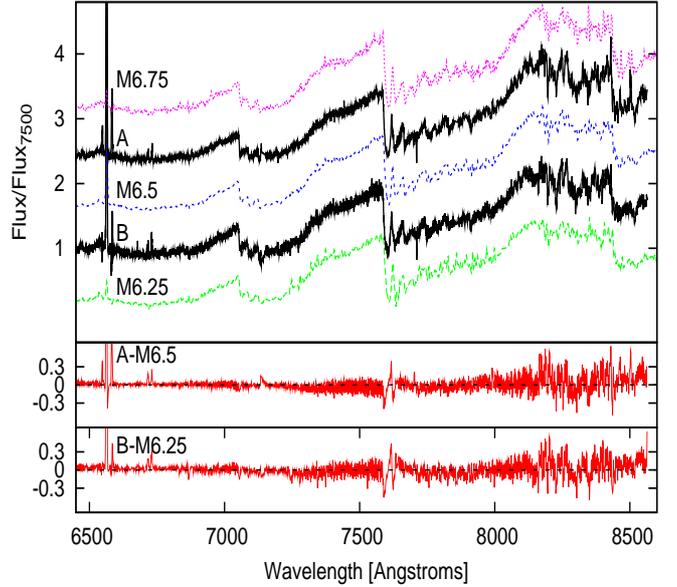}\\
\vspace{1cm}
\caption{Comparison of the GMOS spectra of CHS\,7797 at the two observed phases, i.e. maximum (A) and four days before maximum (B) with Upper Scorpius members spectra from \cite{Lodieu2011}.  The spectra were normalised using the flux at 7500\,$\AA$ and displaced vertically by an arbitrary constant value. The bottom panel shows the subtraction of the spectra of CHS\,7797 at each phase with the best-matched spectra of the UpS members. }
\label{Fig_spectra2}
\end{figure}

A direct comparison of the spectra of CHS\,7797 at the two observed phases and spectra of three members of the $\approx$\,5\,Myr-old Upper Scorpius Association from \cite{Lodieu2011} is shown in Fig.\,\ref{Fig_spectra2}\footnote{The spectral resolution of these spectra is 1100 and therefore, for a proper comparison, we binned our spectra to match its lower resolution. As in the case of CHS\,7797, the Upper Scorpius spectra were flux-calibrated with spectrophotometric standards.}. The spectra are shown with increasing spectral types from bottom to top. At first glance, no differences between the five spectrum are seen. To visualise the differences, we subtracted the observed spectra of CHS\,7797 at both phases with the best-matched spectra of \cite{Lodieu2011} and show the results in the lower panel of Fig.\,\ref{Fig_spectra2}. From this comparison we obtain slightly later spectral types than from the spectral indices described above, namely M6.25 and M6.5 for phase B and A, respectively.

It is worth to mention the uncertainties that affect the spectral type determination. In the case of CHS\,7797 one might expect reddening to give the largest contribution to the error in the spectral type classification. Nevertheless, the lack of any flux-colour correlation at $\lambda$\,$<$\,2$\mu$m indicates that the dust particles in the circumstellar environment of CHS\,7797 are larger than typical dust particles in the interstellar medium, and that we deal with grey extinction. Moreover, as stated above, the carefully selected indices minimise uncertainties that are caused by reddening. An additional source of error, which could explain the differences in the derived SpT at the two observed phases, might result from the atmospheric conditions under which the spectra were obtained. We are aware of a water vapor absorption band centred at 8200$\AA$. Although we lack a measurement of the precipitable water vapor (PWV), weather conditions such as relative humidity ($\approx$\, 35\,\%) and dew-point temperature ($\approx$\, 5$^{\circ}$C), which are related to the amount of PWV, do not differ between the two observing nights, neither does the airmass. In addition, the spectral types derived from the TiO 8465 and VO 7445 indices, which measure the relative fluxes at wavelength ranges that do not include the 8200$\AA$ water absorption band, also differ at the two observed phases\footnote{ We also derived SpT using other spectral indices from \cite{Riddick2007a}, which are not affected by the 8200\,$\AA$ water band and, as expected, they confirm the different spectral types derived at the two distinct phases.}. 

Veiling caused by accretion and spottiness in the photosphere might in principle also affect the spectral classification. Nevertheless, because as in the case of reddening these effects are strongly wavelength dependent, the selection of spectral indices that are defined as flux ratios of very close-by regions of the spectra minimises the uncertainties in the spectral classification. This effect is easily confirmed if we measure the SpT at the two observed phases by means of the commonly used PC3 index \citep{Martin1999}, which is defined as flux ratios of regions that are $\approx$100nm apart, and therefore are not equally affected by extinction or veiling. Based on the PC3 index, CHS\,7797 has the spectral types M5.7 and M4.8 during the bright and faint phase, respectively. This result supports the spectral differences between the two phases found, but 1) these differences are larger and 2) the SpTs derived from the PC3 index are warmer than those obtained from the indices used in this work. These findings can be explained if veiling affects the spectra, in which the bluer part of the spectrum is more affected than the red part and therefore the PC3 index  delivers an earlier spectral type than indices that are not or much less affected by veiling (i.e. the indices defined as flux ratios of close-by spectral regions). This is exactly the case in CHS\,7797, and the PC3 values derived may also indicate that during the faint phase the effect of veiling is stronger than at the brigh phase.

We are therefore confident that the derived spectral types are not strongly affected by any of the previously mentioned effects, in particular because of an appropriate choice of spectral indices. The typical uncertainty for the individual spectral classification method used is 0.25 subclass.

\begin{figure}
\centering
\includegraphics[width=7.2cm,height=6.6cm]{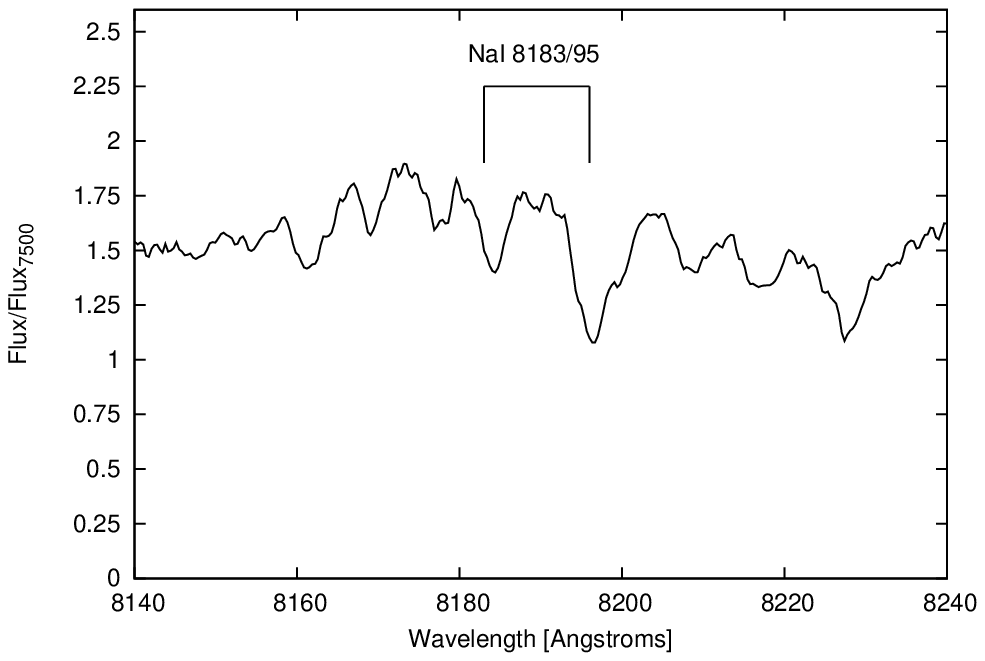}\\
\vspace{1cm}
\includegraphics[width=7.2cm,height=6.6cm]{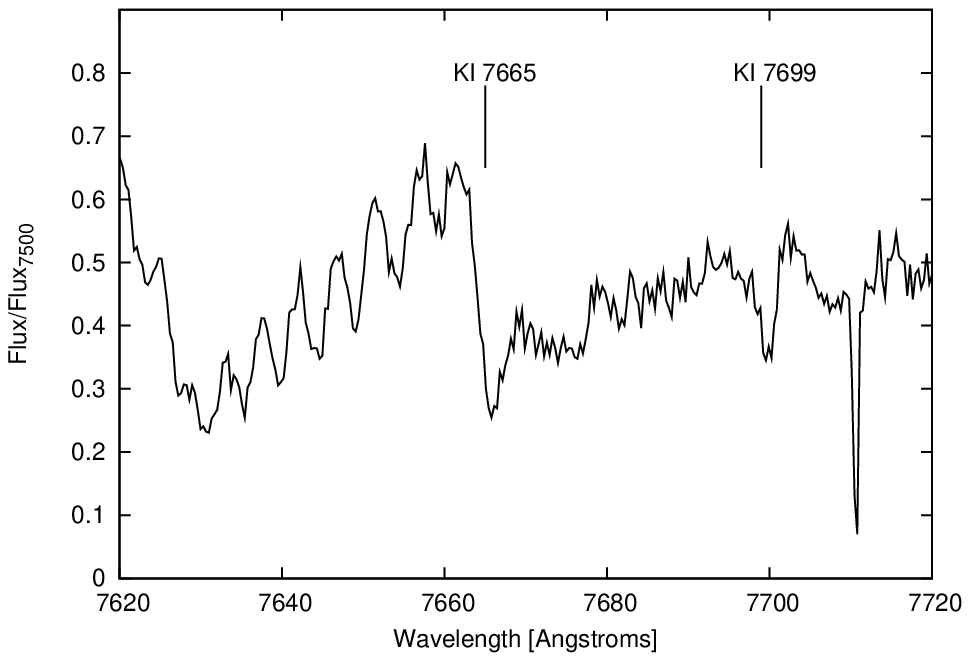}\\
\vspace{1cm}
\includegraphics[width=7.2cm,height=6.6cm]{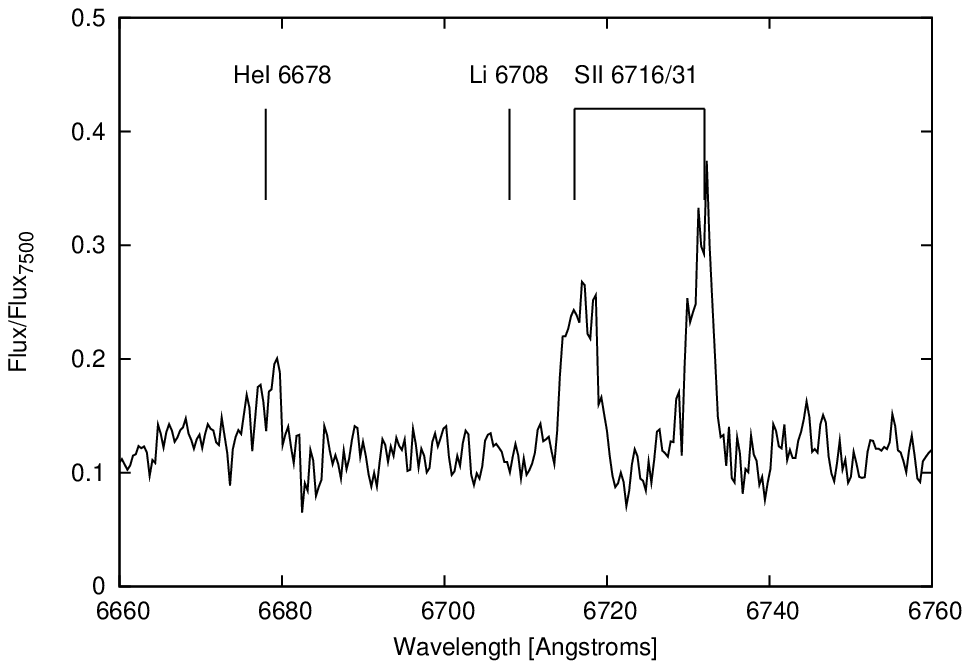}
\vspace{1cm}
\caption{100\,$\AA$ wide zoom of the spectral regions (bright phase) around the gravity-sensitive Na\,I 8183/8195\,$\AA$ doublet \textit{(top)} and the K\,I 7699/7665\,$\AA$ doublet \textit{(middle)}. The \textit{bottom} panel shows the region around the Li\,I resonance doublet at 6708\,$\AA$. }
\label{Fig_lines}
\end{figure}
\section{Luminosity class and membership}\label{luminosity}

The young age of a VLM star or brown dwarf can be inferred from gravity-sensitive spectral features. The radius of an old field VLM star varies only slightly with mass and age, and therefore the surface gravity depends only on the mass. At young ages, low-mass objects have higher effective temperatures and a radius that can be $\approx$\,3 times larger than for an older object of the same SpT \citep{Burrows2001}. These result in significantly lower surface gravities compared to older dwarfs of the same spectral type. Several features in the optical spectrum of M-dwarfs, such as the Na\,I (8183/8195\,$\AA$), K\,I (7665/7699\,$\AA$), and Li\,I (6708\,$\AA$) resonance doublets, can be used to distinguish between young and old dwarfs. Fig.\,3 show the spectral regions around these lines at bright phase. \\
CHS\,7797 is most likely a member of the ONC since it shows signatures of youth such as variability, near-infrared excess, and strong H$\alpha$ emission. In addition, background contaminators are unlikely in this region because of the strong background absorption in the ONC region. Nevertheless, we investigated the membership probability of CHS\,7797 by analysing gravity-sensitive spectral lines to infer its surface gravity and distinguish between a young object and an older field dwarf.   \\                                                                                                                                                                                                                                                                                                                                                                                                                                                                                                                                                                                                                                                                                                                                                                                                                                                                                                                                  
The Na\,I doublet lines (8183/8195\,$\AA$) are commonly used to distinguish between PMS stars and field dwarfs because they are extremely sensitive to gravity and show a tight correlation with SpT \citep[e.g.][]{Martin1999,Riddick2007a}. The absorption is stronger for field dwarfs than for PMS stars and giants. We computed the Na\,I index for CHS\,7797 as defined by \cite{Kirkpatrick1991} and found that CHS\,7797 has a considerably lower Na\,I index value (0.98) than a field dwarf with an SpT between M6 and M7 ($\approx$\,1.2). Our derived Na\,I value of 0.98 agrees with available Na\,I measurements of M6 PMS objects in the ONC \citep[see Fig.\,2 and Fig.\,3 in][]{Riddick2007b}. In addition to the Na\,I index, we computed pseudo-equivalent widths (PEWs) of the Na\,I doublet \textbf{\footnote{We discuss only the PEWs of the doublets and not of the single lines for a proper comparison with the literature.}}by using the IRAF package \textit{splot}. Uncertainties in the PEW calculation were computed based on the full-width at half maximum of the fitted Gaussian as explained in \cite{Kenyon2005}. The derived PEW of the doublet at the two distinct phases are 3.5$\pm$0.2 and 3.3$\pm$0.3\,$\AA$. These low values agree with the PEWs of Na\,I doublet measured in other young stars \citep[e.g][]{Lodieu2011} and are lower than the $\approx$6\,$\AA$ typical equivalent widths measured in M old field dwarfs \citep{Martin1996}. \\
The KI resonance doublet at 7699/7665\,$\AA$ is also a good age indicator. The derived PEWs for the KI resonance doublet are 4.0$\pm$0.2 and 3.4$\pm$0.3\,$\AA$ for the bright and the faint phase observations, respectively. The values agree with the PEW derived by \cite{Lodieu2011} for M6-M7 $\approx$5\,Myr-old stars in Upper Scorpius. Table\,3 shows the PEW of the Na\,I and KI doublets and of each line that constitutes those doublets.\\
Owing to the poor S/N ratio in the region around Li\,I 6708\,$\AA$ we can only give upper limits on the strength of this line. For these upper limits we derive values of $\approx$\,0.3\,$\AA$ in both observed phases. Young M5-M6 objects are found to have a lithium feature with PEW\,$\gtrsim$\,0.3 $\AA$ \citep{Zapatero2002}. Veiling may affect Li\,I feature in CHS\,7797, which would explain the low upper limits. In addition, as can be seen in Fig.\,3, the spectral region around the Li\,I doublet is very noisy and highly affected by the background subtraction of the strong nebular [SII] emission lines as well as from contamination of other lines (e.g. Fe 6707$\AA$). 

\begin{table}[!]
\centering
\renewcommand{\arraystretch}{1.3}  
\caption{Pseudo-equivalent widths (PEWs) in $\AA$ of the gravity sensitive doublets Na\,I and KI at the two observed phases, A (maximum) and B (four days before maximum).}
\label{tab3}
\addtolength{\tabcolsep}{0.0pt}
\begin{tabular}{c|c|c|c|c} 
\hline\hline             
&\multicolumn{2}{|c|}{Phase\,A}&\multicolumn{2}{|c}{Phase\,B}\\
\hline
&PEW&$\Delta$\,PEW&PEW&$\Delta$\,PEW\\
\hline
Na\,I (8183)& 1.0 &0.1& 1.2 & 0.15\\
Na\,I (8195)& 2.5&0.1& 2.1&0.15 \\
Na\,I doublet&3.5&0.2&3.3&0.3\\
K\,I (7665)& 2.5 &0.1& 2.4 & 0.2\\
K\,I (7699)& 1.5&0.1& 1.0&0.1 \\
K\,I doublet&4.0&0.2&3.4&0.3\\
\hline
\end{tabular}
\renewcommand{\footnoterule}{} 
\end{table}

\section{Results and conclusions}

CHS\,7797 is undoubtedly an intriguing very low mass member of the $\approx$\,1\,Myr-old ONC. The spectral type determination based both on spectral indices and on the direct comparison with known young stars in the Upper Scorpius region confirms that CHS\,7797 has a mass that is at the boundary between stars and brown dwarfs. The Na\,I and K\,I gravity-sensitive lines confirm its very young age. Interestingly, we found small differences in the derived SpT at the two distinct phases of the 17.8\,d period. As outlined in the following, these differences, if real\footnote{Note that the differences are only at 1-$\sigma$ level.}, might be a hint for CHS\,7797 being a binary system\textbf{\footnote{We are aware that the best way to prove the binarity of CHS\,7797 is by means of radial velocity studies. Unfortunately, our spectra have a poor S/N. We expect radial velocity differences between the observed phases of $\approx$10\,km/s and an estimated radial velocity uncertainty of the same order. }} in which the two components are of slightly different spectral type. In \cite{Rodriguez-Ledesma2012a} the SED modelling used to characterise CHS\,7797 and its disc suggested the presence of an inner disc cavity, which also argues against the single-object scenario. The best-fitted model predicts an inner disc edge that is located farther from the star than the radius corresponding to a period of 17.8\,d, and therefore no dust structure at this radius can account for the observed obscuration. We consider that the second scenario presented by \cite{Rodriguez-Ledesma2012a} might be the most likely for CHS\,7797, namely that CHS\,7797 is a binary system with an orbital period of 17.8\,d and a circumbinary disc that is slightly inclined with respect to the orbital plane of the binary. In the binary scenario the system is always occulted by the disc but the amount of extinction varies depending on the position of the two stars with respect to the more transparent outer region of the disc. According to this scenario, the slight differences in the two spectra, is an indication for the larger contribution to the total flux from a cooler object during maximum light, which agrees with the slightly later SpT derived during maximum phase compared with the fainter phase. 

To reproduce the differences in both spectra, we performed simple simulations in which different fractions of the flux of both components are considered to contribute to the observed flux at the two distinct phases. We assumed that star B is mostly highly obscured and contributes with less than 25\% of the observed flux at minimum while at maximum light its contribution is zero or up to 5\%. Star A contributes with 90-100\% of its flux during maximum and $\gtrsim$\,50\% during the faint phase. Since for the simulation we used the spectra of the USco members from \cite{Lodieu2011}, we considered that star A is an M6.5 object (i.e. the M6.5 spectra matches the spectra of CHS\,7797 at maximum light best, see Fig.\,\ref{Fig_spectra2}). For star B we considered earlier spectral types from M3.5 to M5.5. Although many combinations are possible, this simple test provides some constraints on the system. Star B has to have an SpT between M4-M5 to reproduce the $\approx$\,0.16 flux excess in the red and star A cannot reduce its flux contribution to less than 60\% during the faint phase, because if it does, the flux excess in the red is stronger than observed independent of star B. A good match with the observed differences is obtained when at maximum phase star A (M6.5) contributes with 95\% of its flux and star B (M5) with 5\%, while at phase 0.75, A contributes with 60\% and B with 30\% of their fluxes. Similar contributions are found if B has a spectral type between M4 and M5. As stated above, this is only an example and only illustrates a possible configuration of the CHS\,7797 system.

To summarise, we presented here spectral classification of the unusual very low mass high-amplitude variable CHS\,7797. Based on four spectral indices, we derived the slightly different spectral types of M6.05\,$\pm$\,0.25 and M5.75\,$\pm$\,0.25 for the observed $phase\,A$ and $phase\,B$, respectively. From the direct comparison with spectra of Upper Scorpius members \citep{Lodieu2011} the derived SpT are M\,6.5 and M\,6.25 for the observed $phase\,A$ and $phase\,B$, respectively. These differences might provide additional hints for the binarity of the system. We also confirmed the young age of CHS\,7797 by means of different measurements of the Na\,I and K\,I resonance doublets, which indicate that CHS\,7797 is most probably a member of the 1\,Myr-old ONC. We could only give upper limits to the Li\,I PEW due to the poor S/N of the spectra in this spectral region. Veiling and spottiness can also affect the PEW of the Li\,I lines, as well as contamination due to the background subtraction from the very strong and the close-by [SII] emission. 

We believe that CHS\,7797 deserves further attention. We plan to continue with photometric and spectroscopic follow-up observations in optical and NIR wavelengths. Higher resolution spectra would allow for radial velocity measurements and would confirm/discard the binary character of CHS\,7797.

\begin{acknowledgements}
We acknowledge the anonymous referee for valuable
comments, which helped to improve the original manuscript. Based on observations obtained with GMOS at the Gemini South Observatory, which is operated by the 
Association of Universities for Research in Astronomy, Inc., under a cooperative agreement 
with the NSF on behalf of the Gemini partnership: the National Science Foundation (United 
States), the Science and Technology Facilities Council (United Kingdom), the 
National Research Council (Canada), CONICYT (Chile), the Australian Research Council (Australia), 
Minist\'{e}rio da Ci\^{e}ncia, Tecnologia e Inova\c{c}\~{a}o (Brazil) 
and Ministerio de Ciencia, Tecnolog\'{i}a e Innovaci\'{o}n Productiva (Argentina). This research was partially funded by the German Research Foundation (DFG) grant HE\,2296/16-1. SB is funded by national program AYA2010-19136 funded by the Spanish ministry of science and innovation.

\end{acknowledgements}
\bibliographystyle{aa}
\bibliography{chs7797a}

\begin{thebibliography}{30}
\expandafter\ifx\csname natexlab\endcsname\relax\def\natexlab#1{#1}\fi

\bibitem[{{Baraffe} {et~al.}(2003){Baraffe}, {Chabrier}, {Allard}, \&
  {Hauschildt}}]{Baraffe2003}
{Baraffe}, I., {Chabrier}, G., {Allard}, F., \& {Hauschildt}, P. 2003, in IAU
  Symposium, Vol. 211, Brown Dwarfs, ed. {E.~Mart{\'{\i}}n}, 41

\bibitem[{{Brice{\~n}o} {et~al.}(2002){Brice{\~n}o}, {Luhman}, {Hartmann},
  {Stauffer}, \& {Kirkpatrick}}]{Briceno2002}
{Brice{\~n}o}, C., {Luhman}, K.~L., {Hartmann}, L., {Stauffer}, J.~R., \&
  {Kirkpatrick}, J.~D. 2002, \apj, 580, 317

\bibitem[{Burrows {et~al.}(2001)Burrows, Hubbard, Lunine, \&
  Liebert}]{Burrows2001}
Burrows, A., Hubbard, W.~B., Lunine, J.~I., \& Liebert, J. 2001, Rev. Mod.
  Phys., 73, 719

\bibitem[{{Chiang} \& {Murray-Clay}(2004)}]{Chiang2004}
{Chiang}, E.~I. \& {Murray-Clay}, R.~A. 2004, \apj, 607, 913

\bibitem[{{Grinin} {et~al.}(2008){Grinin}, {Stempels}, {Gahm}, {Sergeev},
  {Arkharov}, {Barsunova}, \& {Tambovtseva}}]{Grinin2008}
{Grinin}, V., {Stempels}, H.~C., {Gahm}, G.~F., {et~al.} 2008, \aap, 489, 1233

\bibitem[{{Hamilton} {et~al.}(2001){Hamilton}, {Herbst}, {Shih}, \&
  {Ferro}}]{Hamilton2001}
{Hamilton}, C.~M., {Herbst}, W., {Shih}, C., \& {Ferro}, A.~J. 2001, \apjl,
  554, L201

\bibitem[{{Herbst} {et~al.}(2008){Herbst}, {Hamilton}, {Leduc}, {Winn},
  {Johns-Krull}, {Mundt}, \& {Ibrahimov}}]{Herbst2008}
{Herbst}, W., {Hamilton}, C.~M., {Leduc}, K., {et~al.} 2008, \nat, 452, 194

\bibitem[{{Johnson} {et~al.}(2004){Johnson}, {Marcy}, {Hamilton}, {Herbst}, \&
  {Johns-Krull}}]{Johnson2004}
{Johnson}, J.~A., {Marcy}, G.~W., {Hamilton}, C.~M., {Herbst}, W., \&
  {Johns-Krull}, C.~M. 2004, \aj, 128, 1265

\bibitem[{{Kenyon} {et~al.}(2005){Kenyon}, {Jeffries}, {Naylor}, {Oliveira}, \&
  {Maxted}}]{Kenyon2005}
{Kenyon}, M.~J., {Jeffries}, R.~D., {Naylor}, T., {Oliveira}, J.~M., \&
  {Maxted}, P.~F.~L. 2005, \mnras, 356, 89

\bibitem[{{Kirkpatrick} {et~al.}(1991){Kirkpatrick}, {Henry}, \&
  {McCarthy}}]{Kirkpatrick1991}
{Kirkpatrick}, J.~D., {Henry}, T.~J., \& {McCarthy}, Jr., D.~W. 1991, \apjs,
  77, 417

\bibitem[{{Kirkpatrick} {et~al.}(1995){Kirkpatrick}, {Henry}, \&
  {Simons}}]{Kirkpatrick1995}
{Kirkpatrick}, J.~D., {Henry}, T.~J., \& {Simons}, D.~A. 1995, \aj, 109, 797

\bibitem[{{L{\'e}pine} {et~al.}(2003){L{\'e}pine}, {Rich}, \&
  {Shara}}]{Lepine2003}
{L{\'e}pine}, S., {Rich}, R.~M., \& {Shara}, M.~M. 2003, \aj, 125, 1598

\bibitem[{{Lodieu} {et~al.}(2011){Lodieu}, {Dobbie}, \& {Hambly}}]{Lodieu2011}
{Lodieu}, N., {Dobbie}, P.~D., \& {Hambly}, N.~C. 2011, \aap, 527, A24

\bibitem[{{Lodieu} {et~al.}(2009){Lodieu}, {Zapatero Osorio}, {Rebolo},
  {Mart{\'{\i}}n}, \& {Hambly}}]{Lodieu2009}
{Lodieu}, N., {Zapatero Osorio}, M.~R., {Rebolo}, R., {Mart{\'{\i}}n}, E.~L.,
  \& {Hambly}, N.~C. 2009, \aap, 505, 1115

\bibitem[{{Luhman}(2004)}]{Luhman2004}
{Luhman}, K.~L. 2004, \apj, 602, 816

\bibitem[{{Mart{\'{\i}}n} {et~al.}(1999){Mart{\'{\i}}n}, {Delfosse}, {Basri},
  {Goldman}, {Forveille}, \& {Zapatero Osorio}}]{Martin1999}
{Mart{\'{\i}}n}, E.~L., {Delfosse}, X., {Basri}, G., {et~al.} 1999, \aj, 118,
  2466

\bibitem[{{Martin} {et~al.}(1996){Martin}, {Rebolo}, \&
  {Zapatero-Osorio}}]{Martin1996}
{Martin}, E.~L., {Rebolo}, R., \& {Zapatero-Osorio}, M.~R. 1996, \apj, 469, 706

\bibitem[{{Plavchan} {et~al.}(2008){Plavchan}, {Gee}, {Stapelfeldt}, \&
  {Becker}}]{Plavchan2008}
{Plavchan}, P., {Gee}, A.~H., {Stapelfeldt}, K., \& {Becker}, A. 2008, \apjl,
  684, L37

\bibitem[{{Plavchan} {et~al.}(2010){Plavchan}, {Laohakunakorn}, {Seifahrt},
  {Staplefeldt}, \& {Gee}}]{Plavchan2010}
{Plavchan}, P., {Laohakunakorn}, N., {Seifahrt}, A., {Staplefeldt}, K., \&
  {Gee}, A.~H. 2010, in Bulletin of the American Astronomical Society, Vol.~42,
  American Astronomical Society Meeting Abstracts 215, 429.06

\bibitem[{{Reiners}(2009)}]{Reiners2009}
{Reiners}, A. 2009, \apjl, 702, L119

\bibitem[{{Riddick} {et~al.}(2007{\natexlab{a}}){Riddick}, {Roche}, \&
  {Lucas}}]{Riddick2007b}
{Riddick}, F.~C., {Roche}, P.~F., \& {Lucas}, P.~W. 2007{\natexlab{a}}, \mnras,
  381, 1077

\bibitem[{{Riddick} {et~al.}(2007{\natexlab{b}}){Riddick}, {Roche}, \&
  {Lucas}}]{Riddick2007a}
{Riddick}, F.~C., {Roche}, P.~F., \& {Lucas}, P.~W. 2007{\natexlab{b}}, \mnras,
  381, 1067

\bibitem[{{Rodr{\'{\i}}guez-Ledesma} {et~al.}(2009){Rodr{\'{\i}}guez-Ledesma},
  {Mundt}, \& {Eisl{\"o}ffel}}]{Rodriguez-Ledesma2009}
{Rodr{\'{\i}}guez-Ledesma}, M.~V., {Mundt}, R., \& {Eisl{\"o}ffel}, J. 2009,
  \aap, 502, 883

\bibitem[{{Rodr{\'{\i}}guez-Ledesma} {et~al.}(2012){Rodr{\'{\i}}guez-Ledesma},
  {Mundt}, {Ibrahimov}, {Messina}, {Parihar}, {Hessman}, {Alves de Oliveira},
  \& {Herbst}}]{Rodriguez-Ledesma2012a}
{Rodr{\'{\i}}guez-Ledesma}, M.~V., {Mundt}, R., {Ibrahimov}, M., {et~al.} 2012,
  \aap, 544, A112

\bibitem[{{Scholz} \& {Eisl{\"o}ffel}(2004)}]{Scholz2004}
{Scholz}, A. \& {Eisl{\"o}ffel}, J. 2004, \aap, 419, 249

\bibitem[{{Slesnick} {et~al.}(2006){Slesnick}, {Carpenter}, \&
  {Hillenbrand}}]{Slesnick2006}
{Slesnick}, C.~L., {Carpenter}, J.~M., \& {Hillenbrand}, L.~A. 2006, \aj, 131,
  3016

\bibitem[{{Stauffer} {et~al.}(1999){Stauffer}, {Barrado y Navascu{\'e}s},
  {Bouvier}, {Morrison}, {Harding}, {Luhman}, {Stanke}, {McCaughrean},
  {Terndrup}, {Allen}, \& {Assouad}}]{Stauffer1999}
{Stauffer}, J.~R., {Barrado y Navascu{\'e}s}, D., {Bouvier}, J., {et~al.} 1999,
  \apj, 527, 219

\bibitem[{{Winn} {et~al.}(2006){Winn}, {Hamilton}, {Herbst}, {Hoffman},
  {Holman}, {Johnson}, \& {Kuchner}}]{Winn2006}
{Winn}, J.~N., {Hamilton}, C.~M., {Herbst}, W.~J., {et~al.} 2006, \apj, 644,
  510

\bibitem[{{Winn} {et~al.}(2004){Winn}, {Holman}, {Johnson}, {Stanek}, \&
  {Garnavich}}]{Winn2004}
{Winn}, J.~N., {Holman}, M.~J., {Johnson}, J.~A., {Stanek}, K.~Z., \&
  {Garnavich}, P.~M. 2004, \apjl, 603, L45

\bibitem[{{Zapatero Osorio} {et~al.}(2002){Zapatero Osorio}, {B{\'e}jar},
  {Mart{\'{\i}}n}, {Barrado y Navascu{\'e}s}, \& {Rebolo}}]{Zapatero2002}
{Zapatero Osorio}, M.~R., {B{\'e}jar}, V.~J.~S., {Mart{\'{\i}}n}, E.~L.,
  {Barrado y Navascu{\'e}s}, D., \& {Rebolo}, R. 2002, \apjl, 569, L99

\end{thebibliography}
\end{document}